\documentclass[aps,prl,superscriptaddress,nofootinbib]{revtex4}

\usepackage{graphicx}
\usepackage{amssymb}
\usepackage{amsmath}
\usepackage{epsfig}

\usepackage{color}

\begin{document}
\title{Lepton-flavor-violating Higgs decay $h\rightarrow \mu\tau$ and muon anomalous magnetic moment
\\ in a general two Higgs doublet model}

\author{Yuji Omura}
\affiliation{Department of Physics,
          Nagoya University,
      Nagoya, 464-8602, Japan}
\author{Eibun Senaha}
\affiliation{Department of Physics,
          Nagoya University,
      Nagoya, 464-8602, Japan}

\author{Kazuhiro Tobe}
\affiliation{Department of Physics,
          Nagoya University,
      Nagoya, 464-8602, Japan}
\affiliation{Kobayashi-Maskawa Institute for the Origin of Particles and the Universe,
          Nagoya University,
      Nagoya, 464-8602, Japan}

\begin{abstract}
  A two Higgs doublet model (2HDM) is one of minimal extensions of the Standard Model (SM),
  and it is well-known that the general setup predicts the flavor-violating phenomena, mediated
  by neutral Higgs interactions. Recently the CMS collaboration has reported an excess
  of the lepton-flavor-violating Higgs decay in $h\rightarrow \mu\tau$ channel with a
  significance of $2.4$ $\sigma$. We investigate the CMS excess in a general 2HDM with tree-level
  Flavor Changing Neutral Currents (FCNCs), and discuss its impact on the other physical observations.
  Especially, we see that the FCNCs relevant to the excess can enhance the
  neutral Higgs contributions to the muon anomalous magnetic moment, and can resolve the discrepancy between the
  measured value and the SM prediction.
  We also find that the couplings to be consistent with the anomaly of the muon magnetic moment as well as the CMS excess in $h\rightarrow \mu\tau$ predict the sizable rate of $\tau \rightarrow \mu \gamma$, which is within the reach of future B factory.

\end{abstract}
\pacs{14.80.Cp,12.60.Jv}
\maketitle

While a Higgs boson has been discovered at the Large Hadron Collider (LHC) experiment~\cite{Aad:2012tfa,Chatrchyan:2012ufa},
the whole structure of the Higgs sector is
still unknown. Theoretically there is no apparent reason why a Higgs sector
with one Higgs doublet is better than the one with more Higgs doublets. Thus, only the
experimental research will reveal the true answer.

A two Higgs doublet model (2HDM) is a simple extension of the minimal Higgs sector in the SM.
In general, both Higgs doublets couple to fermions, and hence the flavor-changing
Higgs interaction is predicted. This is one of the main differences from the SM.
Recently the CMS collaboration has reported an excess of lepton-flavor-violating Higgs
decay in $h\rightarrow \mu\tau$ mode~\cite{CMS:2014hha,Khachatryan:2015kon}
The SM cannot accommodate such an
excess, however, the general 2HDM~\footnote{Sometimes, it is called the Type III two Higgs doublet model.} can explain the excess, as pointed out
in Refs.~\cite{Sierra:2014nqa,Crivellin:2015mga,deLima:2015pqa}.
\footnote{Multi-Higgs doublet model has been also considered \cite{Heeck:2014qea}. The lepton flavor violating Higgs decays have been investigated before the Higgs discovery \cite{Assamagan:2002kf,Brignole:2003iv,Arganda:2004bz,Kanemura:2004cn,Kanemura:2005hr,Blankenburg:2012ex}. } 
Therefore, it is worth studying it further, and we find that
the $\mu-\tau$ lepton-flavor-violating Higgs interaction can enhance the neutral Higgs contributions
to an anomalous magnetic moment of muon (muon g-2), and hence it can explain the long-standing
anomaly of the muon g-2~\cite{Agashe:2014kda}.

In the general 2HDM, we can always take a basis where only one Higgs doublet gets a vacuum
expectation value (VEV), so that we can parametrize the Higgs doublets as follows;
\begin{eqnarray}
  H_1 =\left(
  \begin{array}{c}
    G^+\\
    \frac{v+\phi_1+iG}{\sqrt{2}}
  \end{array}
  \right),~~~
  H_2=\left(
  \begin{array}{c}
    H^+\\
    \frac{\phi_2+iA}{\sqrt{2}}
  \end{array}
  \right),
\end{eqnarray}
where $G^+$ and $G$ are Nambu-Goldstone bosons, and $H^+$ and $A$ are a charged Higgs boson and a CP-odd
Higgs boson, respectively. CP-even neutral Higgs bosons $\phi_1$ and $\phi_2$ can mix and form mass
eigenstates, $h$ and $H$ ($m_H>m_h$),
\begin{eqnarray}
  \left(
  \begin{array}{c}
    \phi_1\\
    \phi_2
  \end{array}
  \right)=\left(
  \begin{array}{cc}
    \cos\theta_{\beta \alpha} & \sin\theta_{\beta \alpha}\\
    -\sin\theta_{\beta \alpha} & \cos\theta_{\beta \alpha}
  \end{array}
  \right)\left(
  \begin{array}{c}
    H\\
    h
  \end{array}
  \right).
\end{eqnarray}
Here $\theta_{\beta \alpha}$ is the mixing angle.
In mass eigenbasis for the fermions, the Yukawa interactions are expressed as follows;
\begin{align}
  {\cal L}&=-\bar{Q}_L^i H_1 y^i_d d_R^i -\bar{Q}_L^i H_2 \rho^{ij}_d d_R^j \nonumber \\
  &\quad-\bar{Q}_L^i (V^\dagger_{\rm CKM})^{ij}\tilde{H}_1 y^j_u u_R^j -\bar{Q}_L^i (V^\dagger_{\rm CKM})^{ij}\tilde{H}_2
  \rho^{jk}_u u_R^k \nonumber\\
  &\quad-\bar{L}_L^i H_1 y^i_e e_R^i -\bar{L}_L^i H_2 \rho^{ij}_e e_R^j,
\end{align}
where $Q=(V_{\rm CKM}^\dagger u_L,d_L)^T$, $L=(V_{\rm MNS} \nu_L, e_L)^T$,
$V_{\rm CKM} (V_{\rm MNS})$ is the Cabbibo-Kobayashi-Maskawa (Maki-Nakagawa-Sakata) matrix
and the fermions $(f_L,~f_R)$ $(f=u,~d,~e,\nu)$ are mass eigenstates.
$\rho_f^{ij}$ are general 3-by-3 complex matrices and can be sources of 
the Higgs-mediated FCNC processes.
In the following discussions, we do not adopt the so-called Cheng-Sher ansatz~\cite{Cheng:1987rs} for $\rho_f^{ij}$ in order to explore wider parameter space.

In the mass eigenstate of Higgs bosons, the interactions are expressed as
\begin{align}
  {\cal L}&=-\sum_{\phi=h,H,A} y_{\phi i j}\bar{f}_{Li} \phi f_{Rj}
  -\bar{\nu}_{Li} (V_{\rm MNS}^\dagger \rho_e)^{ij}  H^+ e_{Rj}\nonumber \\
  &\quad-\bar{u}_i(V_{\rm CKM}\rho_d P_R-\rho_u^\dagger V_{\rm CKM} P_L)^{ij} H^+d_j+{\rm h.c.},
\end{align}
where
\begin{align}
  y_{hij}&=\frac{m_{f}^i}{v}s_{\beta\alpha}\delta_{ij}+\frac{\rho_{f}^{ij}}{\sqrt{2}} c_{\beta\alpha},~
  y_{Hij}=\frac{m_f^i}{v} c_{\beta \alpha}\delta_{ij}-\frac{\rho_f^{ij}}{\sqrt{2}} s_{\beta\alpha},\nonumber \\
  y_{Aij}&=
    -\frac{i\rho_f^{ij}}{\sqrt{2}}~({\rm for}~f=u),~~
    \frac{i\rho_f^{ij}}{\sqrt{2}}~({\rm for}~f=d,~e),
\end{align}
and $s_{\beta \alpha}=\sin\theta_{\beta\alpha}$ and $c_{\beta \alpha}=\cos\theta_{\beta\alpha}$ are defined.
Note that the SM-like Higgs couplings $y_{hff}$ approach to the SM ones when $c_{\beta\alpha}$ gets closer to zero,
so that the flavor-violating phenomena mediated by the SM-like Higgs boson can be suppressed in this limit.
The current LHC Higgs coupling measurements and search for flavor violation 
suggest the smallness of the mixing parameter $c_{\beta \alpha}$ in this framework.

On the other hand, the CMS collaboration reports that
there is an excess in $h\rightarrow \mu \tau$ process~\cite{CMS:2014hha, Khachatryan:2015kon};
\begin{eqnarray}
  {\rm BR}(h\rightarrow \mu \tau)=(0.84^{+0.39}_{-0.37}) \%,
\end{eqnarray}
where the final state is a sum of $\mu^+\tau^-$ and $\mu^-\tau^+$.
This might be an evidence of a Flavor Changing Neutral Current (FCNC) involving SM-like neutral Higgs,
and, in fact, the flavor-violating coupling $\rho_e$ can accommodate the CMS result in our general 2HDM;
\begin{align}
  {\rm BR}(h \rightarrow \mu \tau)
  &=\frac{c_{\beta\alpha}^2
    (|\rho_e^{\mu \tau}|^2+|\rho_e^{\tau\mu}|^2)m_h}{16\pi\Gamma_h},
\end{align}
where $\Gamma_h$ is a total decay width of Higgs boson $h$ and we adopt $\Gamma_h=4.1$ MeV
in this paper.
In order to explain the excess, the size of the flavor mixing should be as follows;
\begin{align}
  \bar{\rho}^{\mu\tau}&\equiv \sqrt{\frac{|\rho_e^{\mu \tau}|^2+|\rho_e^{\tau\mu}|^2}{2}},
  \nonumber \\
  &\simeq 0.26 \left(\frac{0.01}{c_{\beta\alpha}}  \right)
  \sqrt{\frac{{\rm BR}(h\rightarrow \mu \tau)}{0.84\times 10^{-2}}}.
\end{align}
Even if the Higgs mixing is small ($c_{\beta \alpha}=0.01$),  the O(1) flavor-violating coupling
$\bar{\rho}^{\mu\tau}$ can achieve the CMS excess.

\begin{figure}
  \begin{center}
    {\epsfig{figure=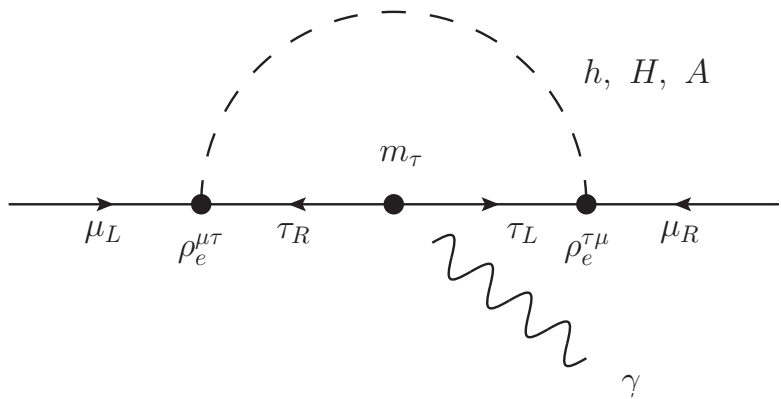,width=0.6\textwidth}}
  \end{center}
  \caption{A Feynman diagram for neutral Higgs boson contributions to the muon g-2.
    A photon is attached somewhere in the charged lepton line.}
  \label{muonG2}
\end{figure}

The next question is what kind of prediction we have, if such a flavor-violating Yukawa coupling exists.
One interesting observable predicted by the FCNC is the muon g-2, where
the discrepancy between the experimental result and the SM prediction is reported.
The CMS excess requires the sizable $\mu-\tau$ flavor violation, so that
it would be possible for the large FCNC to contribute to the muon g-2 through the one-loop diagram 
involving neutral scalars ($h, \, H, \, A$), as we see Fig. \ref{muonG2}.
The extra contributions from $\rho_e^{\mu\tau~(\tau\mu)}$ induce the deviation from the SM
prediction;
%
\begin{equation}
  \delta a_\mu\simeq\frac{m_\mu m_\tau \rho_e^{\mu\tau}\rho_e^{\tau\mu}}{16\pi^2}
  \left[\frac{c^2_{\beta\alpha}(\log\frac{m_h^2}{m_\tau^2}-\frac{3}{2})}{m_h^2}\right.
    \nonumber \quad\left. +\frac{s^2_{\beta\alpha}(\log\frac{m_H^2}{m_\tau^2}-\frac{3}{2})}{m_H^2} 
-\frac{\log\frac{m_A^2}{m_\tau^2}-\frac{3}{2}}{m_A^2}
    \right],
\end{equation}
assuming that $\rho_e^{\mu \tau} \rho_e^{\tau\mu}$ is real, for simplicity.
\footnote{If $\rho_e^{\mu\tau}\rho_e^{\tau\mu}$ is complex, 
the electric dipole moment (EDM) of the muon would be induced. 
The current limit of the muon EDM is $|d_\mu| < 1.8\times 10^{-19}~e\;{\rm cm}$~\cite{Bennett:2008dy},
which is expected to be improved up to $1\times10^{-24}~e\;{\rm cm}$ in the future experiments~\cite{Semertzidis:1999kv,Sato:2009zze}. }
Here we only consider the dominant contributions which are proportional to $\tau$
mass $m_\tau$.\footnote{In general, the other Yukawa couplings $\rho_e$  might contribute
to the muon g-2. Here we have simply assumed that the others are negligible.}
We note that the Yukawa couplings $\rho_e^{\mu\tau~(\tau\mu)}$ generate
an enhancement of $O(m_\tau/m_\mu)$ in the $\delta a_\mu$, where
the $m_\tau$ dependence comes from the internal $\tau$ lepton propagator
in one loop diagram shown in Fig.~\ref{muonG2}.
To maximize a size of the $\delta a_\mu$, while keeping a value of ${\rm BR}(h\rightarrow \mu\tau)$,
$|\rho_e^{\mu \tau}|\sim |\rho_e^{\tau\mu}|$ is preferred.


\begin{figure}
\begin{center}
{\epsfig{figure=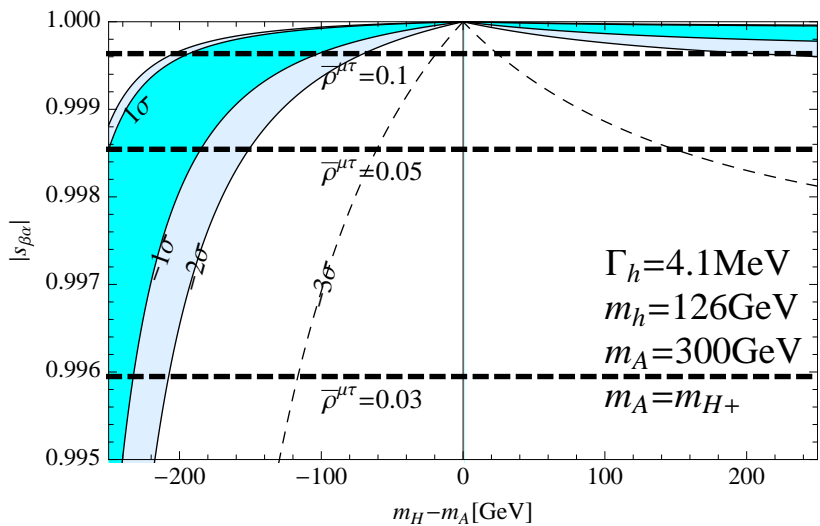,width=0.6\textwidth}}
\end{center}
\vspace{-0.5cm}
\caption{The neutral Higgs contributions to the muon g-2 ($\delta a_\mu$) induced by the lepton flavor violating couplings
  $\rho_e^{\mu\tau(\tau\mu)}$ as functions of $|s_{\beta\alpha}|$ and $m_H-m_A$. Here we assume $\bar{\rho}^{\mu\tau}=\rho_e^{\mu\tau}=
  \pm\rho_e^{\tau \mu}$ where the sign of the $\rho_e^{\tau \mu}$ is fixed to induce the positive contribution
  to $\delta a_\mu$ and
  the value of $\bar{\rho}^{\mu\tau}$ is determined to explain the CMS excess of
  ${\rm BR}(h\rightarrow \mu\tau)$. We have taken $m_A=m_{H^+}=300$ GeV.
  The cyan (light blue) region is the one within $|1\sigma|$ ($|2\sigma|$) range for the muon g-2 anomaly
  with the $1 \sigma$ uncertainty of the CMS $h\rightarrow \mu\tau$ excess. 
  The dashed is $-3 \sigma$ line. The thick dashed lines correspond to
  $\overline{\rho}_{\mu \tau}=0.1, \, 0.05$ and $0.03$ with BR($h \to \mu \tau$)=$0.84 \%$,
  respectively.}
\label{fig2}
\end{figure}

Fig.~\ref{fig2} shows the numerical result of $\delta a_\mu$ induced by the lepton-flavor-violating couplings
$\rho_e^{\mu\tau~(\tau\mu)}$ as functions of $|s_{\beta\alpha}|$ and a mass difference between $H$ and $A$, $m_H-m_A$.
Here we have assumed $\bar{\rho}^{\mu\tau}=\rho^{\mu\tau}_e=\pm \rho_e^{\tau\mu}$ where the sign of $\rho_e^{\tau\mu}$ is
chosen to realize the positive contribution to $\delta a_\mu$ and the value of $\bar{\rho}_e^{\mu\tau}$ is determined
to explain the CMS excess of ${\rm BR}(h\rightarrow \mu\tau)$. We have taken $m_A=m_{H^+}=300$ GeV.
In the cyan (light blue) region of Fig.~\ref{fig2}, the anomaly of the muon g-2 can be explained within
$|1\sigma|$  ($|2\sigma|$)  with the $1\sigma$ uncertainty of the CMS $h\rightarrow \mu\tau$ excess. 
The $-3\sigma$ line for the muon g-2 anomaly is also shown. Here we adopt the value of the muon g-2 anomaly
from Ref.~\cite{Hagiwara:2011af},
$\delta a_\mu=(26.1\pm 8.0)\times 10^{-10}$.
In Fig.~\ref{fig2}, the thick dashed lines correspond to $\bar{\rho}^{\mu\tau}=0.1,~0.05$
and $0.03$ with ${\rm BR}(h\rightarrow \mu\tau)=0.84\%$, respectively.

In order to explain the anomaly of the muon g-2, the Higgs mixing parameter $|s_{\beta\alpha}|$
should be close to one, which is consistent with the current Higgs coupling measurements at the
LHC experiment.
Note that the non-degeneracy among neutral Higgs bosons induces the larger $\delta a_\mu$.
Although the non-degeneracy also generates the extra contributions to Peskin-Takeuchi's T-parameter \cite{Marciano:1990dp,Kennedy:1991sn,Altarelli:1990zd,peskin},
we have found that the small Higgs mixing parameter $c_{\beta\alpha}$ suppresses the
extra contributions in the current scenario when $m_A$ is very close to $m_{H^+}$.

As pointed out in Refs.~\cite{Davidson:2010xv,Sierra:2014nqa}, the 
Yukawa couplings $\rho_e^{\mu\tau~(\tau\mu)}$ would also induce significant contributions to
$\tau \rightarrow \mu \gamma$
process.
The amplitude of $\tau \rightarrow \mu \gamma$ process is parametrized by
\begin{eqnarray}
T=e \epsilon^{\alpha *}\bar{u}_\mu m_\tau i\sigma_{\alpha \beta} q^\beta (A_L P_L+A_R P_R)u_\tau,
\end{eqnarray}
where $P_{R,~L}(=(1\pm \gamma_5)/2)$ are chirality projection operators, and $e,~\epsilon^\alpha,~q$
and $u_f$ are the electric charge, a photon polarization vector, a photon momentum, and a spinor of
the fermion $f$, respectively. The branching ratio is given by
\begin{eqnarray}
  \frac{{\rm BR}(\tau \rightarrow \mu \gamma)}{{\rm BR}(\tau \rightarrow \mu \bar{\nu}_\mu \nu_\tau)}
  =\frac{48\pi^3\alpha\left(|A_L|^2+|A_R|^2\right)}
  {G_F^2},
\end{eqnarray}
where $\alpha$ and $G_F$ are the fine structure constant and Fermi constant, respectively.
The lepton-flavor-violating Higgs contributions to $A_L$ and $A_R$ are given by
\begin{align}
  A_{L,~R}&=\sum_{\phi=h,~H,~A,~H^-} A_{L,~R}^{\phi},\\
  A_L^{\phi}&= \frac{y_{\phi\tau \tau}y_{\phi\tau\mu}}{16\pi^2 m_\phi^2}\left(\log\frac{m_\phi^2}{m_\tau^2}-\frac{4}{3}
  \right),~~(\phi=h,~H)\nonumber \\
  A_L^{A}&= \frac{y_{A\tau \tau}y_{A\tau\mu}}{16\pi^2 m_A^2}\left(\log\frac{m_A^2}{m_\tau^2}-\frac{5}{3}
  \right),\nonumber \\
  A_R^{\phi}&=A_L^{\phi}|_{y_{\phi \tau\mu}\rightarrow y_{\phi \mu\tau}},~~(\phi=h,~H,~A),\nonumber \\
  A_L^{H^-}&= -\frac{(\rho_e^\dagger \rho_e)^{\mu \tau}}{192\pi^2 m_{H^-}^2},~~
  A_R^{H^-}=0,
\end{align}
where $A_{L,~R}^{\phi}~(\phi=h,~H,~A,~H^-)$ are the $\phi$ contributions at the one loop level.
We also include Barr-Zee-type two-loop contributions
to $A_{R,~L}$ in the numerical analysis, as studied in
Refs.~\cite{Chang:1993kw,Sierra:2014nqa,Davidson:2010xv}.
\footnote{We have found a disagreement between our expression of the one loop contributions $A_{R,~L}^\phi$
and one given in Refs.~\cite{Sierra:2014nqa,Davidson:2010xv}, and our relative sign between the one and
two loop contributions differs from one in Refs.~\cite{Sierra:2014nqa,Davidson:2010xv}.}
\begin{figure}
  \begin{center}
{\epsfig{figure=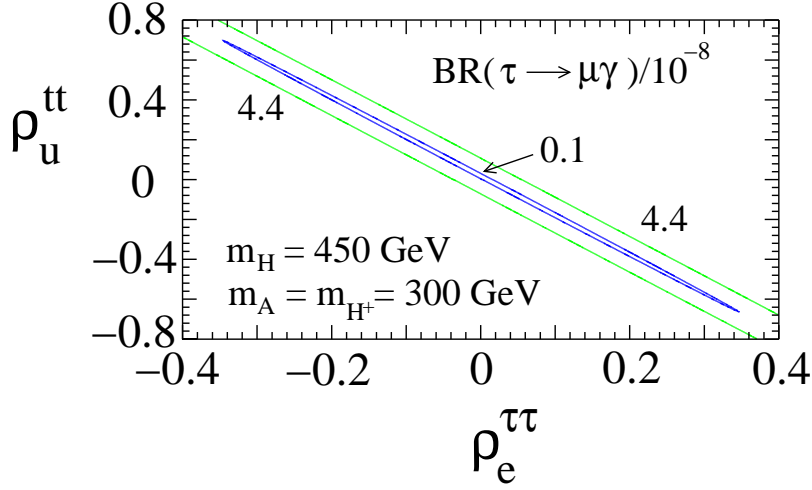,width=0.6\textwidth}}    
  \end{center}
  \caption{Branching ratio of $\tau\rightarrow \mu\gamma$ is shown as a function of $\rho_e^{\tau\tau}$ and $\rho_u^{tt}$.
    It is assumed that $m_H=450$ GeV and $m_A=m_{H^+}=300$ GeV, $s_{\beta\alpha}=0.9999$ and $\rho_e^{\mu\tau}=-\rho_e^{\tau\mu}$
    whose values are determined to realize ${\rm BR}(h\rightarrow \mu\tau)=0.84\%$.
    The lines for ${\rm BR}(\tau\rightarrow \mu\gamma)=4.4\times 10^{-8}$ (current limit) and
    $1\times 10^{-9}$ are shown. For this parameter set, the predicted value of $\delta a_\mu$ is
    $2.1\times 10^{-9}$.}
  \label{BR_tmg}
\end{figure}
When we assume non-zero $\rho_e^{\mu\tau(\tau\mu)}$ as suggested by the CMS excess in $h\rightarrow \tau\mu$,
but other $\rho_f$ couplings are negligibly small, the predicted branching ratio of $\tau \rightarrow \mu \gamma$ is 
smaller than the current experimental limit (${\rm BR}(\tau \rightarrow \mu \gamma)<4.4\times 10^{-8}$ at the 90\% CL.~\cite{Hayasaka:2007vc,Aubert:2009ag}),
however, it would be within a reach of the future B-factory. If unknown Yukawa couplings $\rho_f$ other than $\rho_e^{\mu\tau (\tau\mu)}$ are non-zero, the branching ratio can be significantly increased.

Fig.~\ref{BR_tmg} shows the branching ratio of $\tau \rightarrow \mu \gamma$ as functions of $\rho_e^{\tau\tau}$ and $\rho_u^{t t}$ in the presence of
the non-zero $\rho_e^{\mu\tau(\tau\mu)}$. 
Note that $\rho_u^{tt}$ appears in the Barr-Zee diagrams.
Here we have assumed that other $\rho_f$ Yukawa couplings are negligible, and $m_H=450$ GeV, $m_A=m_{H^+}=300$ GeV
and $s_{\beta\alpha}=0.9999$ are given. We choose $\rho_e^{\mu\tau}=-\rho_e^{\tau\mu}$ to achieve the positive contribution to $\delta a_\mu$ and the values of $\rho_e^{\mu\tau(\tau\mu)}$ are
determined to explain the CMS excess, ${\rm BR}(h\rightarrow \mu\tau)=0.84\%$. In Fig.~\ref{BR_tmg}, the line for the current experimental limit
${\rm BR}(\tau\rightarrow \mu\gamma)=4.4\times 10^{-8}$~\cite{Hayasaka:2007vc,Aubert:2009ag} is shown.
One sees that the limit strongly constrains $\rho_e^{\tau\tau}$ and $\rho_u^{tt}$, 
however, they can still be of $O(1)$ if the signs of them are opposite,
which is due to a cancellation between the one- and two-loop contributions.
The line for a future reference ${\rm BR}(\tau\rightarrow \mu\gamma)=1\times 10^{-9}$~\cite{Aushev:2010bq}
is also shown.
As one can see from Fig.~\ref{BR_tmg}, even if $\rho^{tt}_u=\rho^{\tau\tau}_e=0$ is satisfied, the branching ratio can be as large as $10^{-9}$.
The future improvement on the search for $\tau\rightarrow \mu\gamma$ at the level of $10^{-9}$ will be crucial to test this scenario.
In passing, the nonzero $\rho_u^{tt}$ can contribute to $\delta a_\mu$ via the Barr-Zee
diagrams. However, it is found that its effect is subdominant.

For other tau decay modes~\cite{OST}, non-zero $\rho_e^{\mu \tau (\tau\mu)}$ couplings induce a correction to
$\tau \rightarrow \mu\nu\bar{\nu}$ mode.  We find that the corrction is of O($10^{-5}-10^{-3}$) for the parameter
space where the muon g-2 can be explained, and it is consistent with the current experimental results.
For $\tau \rightarrow \mu ll$ $(l=\mu, \, e)$,
the non-zero branching ratios are predicted even if only $\rho_e^{\mu \tau (\tau\mu)}$ are non-zero. 
The predicted rate, however, is well below the current experimental limit. The rate strongly depends
on $\rho_e^{ll}$, and the current limit is setting a strong constraint as $\rho_e^{ll} \lesssim 0.01$
for the parameter set studied in Fig.~\ref{BR_tmg}. The future improvement of the sensitivity will be very important.
\footnote{The other flavor violating $\tau$ decay such as $\tau \to \mu \eta $ may also give stringent constraints on the  Yukawa couplings in quark sector. For instance, we estimate the upper bound on $\rho^{ss}_d$ from the bound on ${\rm BR}(\tau \to \mu \eta)$: $\rho^{ss}_d\lesssim 0.01$ in the case with $m_A=350$ GeV. This is almost the same order as the one of $\rho^{\mu \mu}_e$.}


A general 2HDM may be also responsible for discrepancies in $B\rightarrow D\tau\nu$, $B\rightarrow D^*\tau\nu$
and $B\rightarrow \tau\nu$ processes as studied in Ref.~\cite{Crivellin:2012ye}. The couplings $\rho_e^{\mu\tau (\tau\mu)}$ can contribute
to $B\rightarrow D\tau\nu$, $B\rightarrow D^*\tau\nu$ and $B\rightarrow \tau\nu$ via a charged Higgs mediation if Yukawa
couplings $\rho_u$ relevant to these processes are sizable. However, since the sizable
contribution to the muon g-2 requires $\rho_e^{\mu\tau}\sim \rho_e^{\tau\mu}$, they also induce the significant contributions to
$B\rightarrow D \mu\nu$, $B\rightarrow D^*\mu\nu$ and $B\rightarrow \mu\nu$ processes, so that it would be difficult to
explain these discrepancies, and the relevant Yukawa couplings $\rho_{u(d)}$ should be negligible in our scenario.

In order to explain the muon g-2 anomaly, the relatively light extra Higgs bosons $A$, $H$, and $H^\pm$ are
required. They will be expected to be produced at the LHC experiment. The production via quark Yukawa
couplings $\rho_{u,d}$ will be possible and important. Furthermore, in the presence of the sizable $\rho_u^{tt}$,
the gluon fusion production process for $A$ and $H$ would be dominant.
However, it is difficult to predict the production cross section without the detail knowledge of the
Yukawa couplings $\rho_{e,u,d}$. On the other hand,
the production via weak interaction such as $q\bar{q}'\rightarrow W^{\pm*}\rightarrow AH^\pm$
is less model-dependent as discussed in Ref.~\cite{Cao:2003tr}.
The current LHC experimental data would put constraints on various unknown Yukawa couplings $\rho_f$.
The detail study will be worth probing this scenario and we will report it in a forthcoming paper~\cite{OST}.

In conclusion, the CMS experiment has reported the excess in $h\rightarrow \mu\tau$.
Although the definitive statement cannot be made until 
the statistical significance of this excess becomes higher 
and the ATLAS collaboration also confirms it,
this might be a hint for new physics.
The general 2HDM can easily accommodate the excess, which can be induced by the $\mu-\tau$ lepton-flavor-violating couplings. We have found that the $\mu-\tau$ flavor violation can significantly enhance the neutral Higgs contributions to the muon g-2, and hence it
can explain the anomaly of the muon g-2. 
In the parameter region where both anomalies for $h \rightarrow \mu\tau$ and
the muon g-2 can be solved, the branching ratio of  $\tau \rightarrow \mu \gamma$ can be sizable
and the search at the future B factory would be crucial to test this scenario.
Since the flavor structure of new Yukawa couplings $\rho_{e,u,d}$ is unknown, the further experimental and theoretical studies would be important to reveal the scenario. This will be just a beginning of many of new phenomena beyond the SM.

\vspace{1cm}

This work was supported in part by Grants-in-Aid for Scientific Research from the Ministry of Education,
Culture, Sports, Science, and Technology (MEXT), Japan (No.22224003 for K.T. and No. 23104011 for Y.O.)
and Japan Society for Promotion of Science (JSPS) (No.26104705 for K.T.)
\bibliographystyle{apsrev}

\end{document}